\documentstyle[11pt,epsfig,a4,twoside]{article}
\def\B{\mathop{\rm B}}
\def\c{\mathop{\rm c}}
\def\cl{\mathop{\rm cl}}

\def\df{\mathop{\rm d}}
\def\dec{\mathop{\rm dec}}

\def\eq{\mathop{\rm eq}}
\def\etal{{\em et al.}\,}
\def\F{\mathop{\rm F}}
\def\fr{{\rm fr}}
\def\L{\mathop{\rm L}}
\def\m{\mathop{\rm m}}

\def\N{\mathop{\rm N}}

\def\pr{{\rm p}}

\def\rec{\mathop{\rm rec}}

\def\with{\mathop{\rm with}}
\def\lessim{\mathbin{\;\raise1pt\hbox{$<$}\kern-8pt\lower3pt\hbox{$\sim$}\;}
}
\def\gtrsim{\mathbin{\;\raise1pt\hbox{$>$}\kern-8pt\lower3pt\hbox{$\sim$}\;}}
\def\95cl{\mathop{95 \%\ \rm c.l.}}
\def\cm{\mathop{\rm cm}}
\def\dK{\mathop{\rm K}}
\def\eV{\mathop{\rm eV}}
\def\GeV{\mathop{\rm GeV}}
\def\gm{\mathop{\rm gm}}
\def\Gyr{\mathop{\rm Gyr}}
\def\keV{\mathop{\rm keV}}
\def\km{\mathop{\rm km}}
\def\kpc{\mathop{\rm kpc}}
\def\MeV{\mathop{\rm MeV}}
\def\Mpc{\mathop{\rm Mpc}}
\def\sec{\mathop{\rm sec}}
\def\hy{\mathop{\rm H}}
\def\h2{\mathop{\rm D}}
\def\hel3{^3{\rm He}}
\def\he4{^4{\rm He}}
\def\li7{^7{\rm Li}}
\newcommand{\case}[2]{{\textstyle\frac{#1}{#2}}}
\begin{document}
\title{Neutrinos from the Big Bang}
\author{{\bf Subir Sarkar} \\ 
   {\small\sl Department of Physics, University of Oxford,
     1 Keble Road, Oxford OX1 3NP, UK}}
\date{}
\maketitle
\begin{abstract}
The standard Big Bang cosmology predicts the existence of an, as yet
undetected, relic neutrino background, similar to the photons observed
in the cosmic microwave background. If neutrinos have mass, then such
relic neutrinos are a natural candidate for the dark matter of the
universe, and indeed were the first particles to be proposed for this
role. This possibility has however been increasingly constrained by
cosmological considerations, particularly of large-scale structure
formation, thus yielding stringent bounds on neutrino masses, which
have yet to be matched by laboratory experiments. Another probe of
relic neutrinos is primordial nucleosynthesis which is sensitive to
the number of neutrino types (including possible sterile species) as
well to any lepton asymmetry. Combining such arguments with the
experimental finding that neutrino mixing angles are large, excludes
the possibility of a large asymmetry and disfavours new neutrinos
beyond those now known.
\end{abstract}
\thispagestyle{empty}
\pagestyle{plain}
\section{Relic neutrinos and the cosmological mass bound}

Several years before neutrinos had even been experimentally detected,
Alpher, Follin \& Hermann \cite{afh53} noted that they would have
been in thermal equilibrium in the early universe ``\ldots through
interactions with mesons'' at temperatures above $5\MeV$; below this
temperature the neutrinos ``\ldots freeze-in and continue to expand
and cool adiabatically as would a pure radiation gas''. These authors
also observed that the subsequent annihilation of $e^\pm$ pairs would
heat the photons but not the decoupled neutrinos, so by conservation
of entropy $T_\nu/T$ would decrease from its high temperature value of
unity, down to $(4/11)^{1/3}$ at $T\ll\,m_e$ \cite{afh53}.

(Thus the present density of {\em massless} relic neutrinos
(neglecting a possible chemical potential i.e. lepton--antilepton
asymmetry) would be
\begin{equation}
\label{nnu}
 \frac{n_\nu}{n_\gamma} = \left(\frac{T_\nu}{T}\right)^3
                          \left(\frac{n_\nu}{n_\gamma}\right)_{T=T_{\dec}}
                      = \frac{4}{11} \left(\frac{3}{4}\frac{g_\nu}{2}\right) ,
\end{equation} 
where $g_\nu=2$ corresponds to left-handed neutrinos and
right-handed antineutrinos, and the factor 3/4 reflects Fermi versus
Bose statistics. This would also be true for massive neutrinos if the
neutrinos are relativistic at decoupling i.e. for
$m_\nu\ll\,T_{\dec}$. For a present cosmic microwave background (CMB)
blackbody temperature $T_0=2.725\pm0.002\dK$ \cite{Hagiwara:fs}, the
abundance per flavour should then be
$\case{3}{11}\times\case{2\zeta(3)}{\pi^2}T_0^3\simeq111.9\cm^{-3}$.
As long as they remain relativistic, the neutrinos would retain a
Fermi-Dirac distribution with phase-space density
\begin{equation}
\label{fnu}
 f_{\nu} = \frac{g_\nu}{(2\pi)^3} 
           \left[\exp\left(\frac{p}{T_\nu}\right)+1\right]^{-1} ,
\end{equation}
since the momentum and temperature would redshift identically.)

Subsequently, Chiu \& Morrison \cite{cm61} found the rate for
$e^+e^-\rightleftharpoons\nu_e\bar{\nu_e}$ in a plasma to be
$\Gamma_{\nu}\approx\,G_{\F}^2T^5$ for the universal Fermi interaction
and Zel'dovich \cite{z6566} equated this to the Hubble expansion rate
in the radiation-dominated era,
\begin{equation}
\label{H}
 H = \sqrt{\frac{8\pi G_{\N} \rho}{3}} ,\quad {\with}\ 
     \rho = \frac{\pi^2}{30} g_* T^4 ,
\end{equation}
where $g_*$ counts the relativistic degrees of freedom, to obtain the
`decoupling' temperature below which the neutrinos expand freely
without further interactions as
$T_{\dec}(\nu_e)\simeq2\MeV$. (Neutral currents were then unknown so
$T_{\dec}(\nu_\mu)$ was estimated from the reaction
$\mu\rightleftharpoons\,e\bar{\nu_e}\nu_\mu$ to be $12\MeV$. Later De
Graaf \cite{dg70} noted that these would keep $\nu_\mu$'s coupled to
the plasma down to the same temperature as $\nu_e$'s.\footnote{In fact
$T_{\dec}(\nu_\mu,\nu_\tau)\simeq3.5\MeV$ while
$T_{\dec}(\nu_e)\simeq2.3\MeV$ because of the additional charged
current reaction \cite{Dicus:bz,Enqvist:gx}. Actually decoupling is
not an instantaneous process so the neutrinos are slightly heated by
the subsequent $e^+e^-$ annihilation, increasing the number density
(\ref{nnu}) by $\approx1\%$ \cite{Dodelson:1992km,Hannestad:1995rs}.})
However Zel'dovich \cite{z6566} and Chiu \cite{c66} concluded that
relic neutrinos, although nearly as numerous as the blackbody photons,
cannot make an important contribution to the cosmological energy
density since they are probably massless.

Interestingly enough, some years earlier Pontecorvo \& Smorodinski
\cite{ps61} had discussed the bounds set on the cosmological energy
density of MeV energy neutrinos (created e.g. by large-scale
matter-antimatter annihilation) using data from the Reines--Cowan and
Davis experiments.\footnote{These authors \cite{ps61} were also the
first to suggest searching for high energy neutrinos by looking for
upward going muons in underground experiments, the basis for today's
neutrino telescopes.}) Not surprisingly these bounds were rather weak
so these authors stated somewhat prophetically that ``\ldots it is not
possible to exclude a priori the possibility that the neutrino and
antineutrino energy density in the Universe is comparable to or larger
than the average energy density contained in the proton rest
mass''. Zel'dovich \& Smorodinski \cite{zs61} noted that better
bounds can be set by the limit on the cosmological energy density in
any form of matter $\rho_{\m}~(\equiv\Omega_{\m}\rho_{\c}$) following
from the observed present expansion rate $H_0$ and age $t_0$ of the
universe.\footnote{The critical density is
$\rho_{\c}=3H_0^2/8\pi\,G_{\N}\simeq1.879\times10^{-29}h^2\gm\,\cm^{-3}$
where the Hubble parameter $h\equiv\,H_0/100\km\,\sec^{-1}\Mpc^{-1}$,
so the present Hubble age is $H_0^{-1}=9.778\,h^{-1}\Gyr$.} Of course
they were still discussing {\em massless} neutrinos. Weinberg
\cite{w62} even speculated whether a degenerate sea of relic neutrinos
can saturate the cosmological energy density bound and noted that such
a sea may be detectable by searching for (scattering) events beyond
the end-point of the Kurie plot in $\beta$-decay experiments.

Several years later, Gershte\u{\i}n \& Zel'dovich \cite{Gershtein:gg}
made the connection that if relic neutrinos are massive, then a bound
on the mass follows from simply requiring that
\begin{equation}
 m_\nu\,n_\nu < \rho_{\m} .
\end{equation}
Using the general relativistic constraint
$\Omega\,t_0^2\,H_0^2<(\pi/2)^2$, they derived
$\rho_{\m}<2\times10^{-28}\gm\cm^{-3}$ (just assuming that
$t_0>5\Gyr$, i.e. that the universe is older than the Earth) and
inferred that $m_{\nu_e},m_{\nu_\mu}<400$~eV for a present photon
temperature of $3~\dK$. Their calculation of the relic neutrino
abundance was rather approximate --- they adopted $g_\nu=4$
i.e. assumed massive neutrinos to be Dirac particles with fully
populated right-handed (RH) states (even though they commented {\em en
passim} that according to the $V-A$ theory such states are
non-interacting and would thus not be in equilibrium at $T_{\dec}$)
and moreover they did not allow for the decrease in the neutrino
temperature relative to photons due to $e^+e^-$
annihilation. Nevertheless their bound was competitive with the best
laboratory bound on $m_{\nu_e}$ and $10^4$ times better than that on
$m_{\nu_\mu}$, demonstrating the sensitivity (if not the precision!) of
cosmological arguments.

A better bound of $m_{\nu_\mu}<130$~eV was quoted by Marx \& Szalay
\cite{ms72} who numerically integrated the cosmological Friedmann
equation from $\nu_\mu$ decoupling down to the present epoch, subject
to the condition $t_0>4.5\Gyr$. Independently Cowsik \& McClleland
\cite{Cowsik:gh} used direct limits on $\Omega_{\m}$ and $h$ to obtain
an even more restrictive bound of $m_\nu<8$~eV, assuming that
$m_\nu=m_{\nu_e}=m_{\nu_\mu}$; however they too assumed that $T_\nu=T$
and that RH states were fully populated. As Shapiro, Teukolsky \& Wasserman
\cite{Shapiro:rr} first emphasized, even if massive neutrinos are
Dirac rather than Majorana, the RH states have no gauge interactions
so should have decoupled much earlier than the left-handed ones. Then
subsequent entropy generation by massive particle annihilations would
have diluted their relic abundance to a negligible
level.\footnote{Although spin-flip scattering (at a rate
$\propto(m_\nu/T)^2$) {\em can} generate RH states, this can be
neglected for $m_\nu\ll1\MeV$. Even if RH neutrinos have new
(superweak) interactions, their relic abundance can be no more than
$\sim10\%$ of LH neutrinos so the bound on their mass (\ref{mnu}) is
relaxed to $\sim1$~keV \cite{Olive:1981ak}. Even if the
(post-inflationary) universe was not hot enough to bring such
interactions into equilibrium, a cosmologically interesting abundance
can still be generated if the RH states have small mixings with LH states
\cite{Dolgov:2000ew}.} Thus we arrive at the modern version of the
`Gershte\u{\i}n-Zel'dovich' bound' \cite{Bernstein:iy}: the
conservative limits $t_0>10\Gyr$ and $h>0.4$ imply $\Omega_{\m}h^2<1$
i.e. $\rho_{\m}<10.54\keV\cm^{-3}$ \cite{Kolb:vq}; combining this with
the relic neutrino number density, which is $\sim1\%$ larger
\cite{Hannestad:1995rs} than in eq.(\ref{nnu}), gives:\footnote{If
neutrinos are non-relativistic at decoupling, then they drop out of
chemical equilibrium with an abundance inversely proportional to their
self-annihilation cross-section so
$\Omega_{\nu}h^2\approx\,(m_\nu/2\GeV)^{-2}$ for $m_\nu\ll\,m_Z$
\cite{Lee:1977ua,Vysotsky:pe}. Thus neutrinos with a mass of ${\cal
O}(\GeV)$ can also account for the dark matter; however LEP has ruled
out such (possible 4th generation) neutrinos upto a mass $\sim\,m_Z/2$
\cite{Hagiwara:fs}. (Conversely $\Omega_{\nu}h^2>1$ for the mass range
$\sim100\eV-2\GeV$, which is thus cosmologically {\em forbidden} for
any stable neutrino having only electroweak interactions
\cite{Sato:1977ye,Dicus:1977qy}.) For heavier masses upto the highest
plausible value of ${\cal O}$(TeV), the relic abundance decreases
steadily due to the increasing annihilation cross-section and remains
cosmologically uninteresting \cite{Enqvist:1988we}.}
\begin{equation}
\label{mnu}
 \Omega_\nu\,h^2 =
 \sum_i \left(\frac{m_{\nu_i}}{93\eV}\right)\left(\frac{g_{\nu_i}}{2}\right)
 < 1 .
\end{equation} 
This is a rather conservative bound since galaxy surveys indicate much
tighter constraints on the total amount of gravitating (dark) matter
in the universe, e.g. the observed `redshift space distortion'
\cite{Hawkins:2002sg} suggests $\Omega_{\m}\approx0.3$, averaged over
a volume extending several hundred Mpc. Together with the Hubble Key
Project determination of $h=0.72\pm0.08$ \cite{Freedman:2000cf}, this
implies that the sum of all neutrino masses cannot exceed about 15~eV.

Although this has historically been the most restrictive constraint on
neutrino masses, it is no longer competetive with the direct
laboratory bound on the electron neutrino mass from the Mainz and
Troitsk tritium $\beta$-decay experiments \cite{Weinheimer:2002rs}:
\begin{equation}
\label{mnulab}
 m_\nu < 2.2~\eV (\95cl).
\end{equation}
Although the kinematic mass limits on the other neutrino flavours are
much weaker (viz. $m_{\nu_\mu}<190$~keV, $m_{\nu_\tau}<18.2$~MeV
\cite{Hagiwara:fs}), the bound above now applies in fact to {\em all}
eigenstates \cite{Barger:1998kz} given the rather small
mass-differences indicated by the oscillation interpretation of the
Solar ($\Delta\,m^2\simeq7\times10^{-5}\eV^2$) and atmospheric
($\Delta\,m^2\simeq3\times10^{-3}\eV^2$) neutrino anomalies
\cite{Gonzalez-Garcia:2002dz}. This means that the laboratory limit on
the sum of neutrino masses (for comparison with the new cosmological
bounds to be discussed) is presently 6.6 eV; the sensitivity of
planned future experiments is at the eV level
\cite{Osipowicz:2001sq}. Moreover relic neutrinos contribute at least
$\Omega_\nu\,h^2\sim0.07/93\sim8\times10^{-4}$ to the cosmic budget
(assuming a mass hierarchy), nearly as much as visible baryons
\cite{Fukugita:1997bi}.

\section{Neutrinos as the galactic `missing mass'}

The cosmological bound (\ref{mnu}) assumes conservatively that
neutrinos constitute {\em all} of the (dark) matter permitted by the
dynamics of the universal Hubble expansion. Further constraints must
be satisfied if they are to cluster on a specified scale
(e.g. galactic halos or galaxy clusters) and provide the dark matter
whose presence is inferred from dynamical measurements. Cowsik \&
McClleland \cite{cm72b} were the first to suggest that neutrinos with
a mass of a few eV could naturally be the `missing mass' in clusters
of galaxies. This follows from the relation
$m_\nu^8\simeq1/G_{\N}^3r_{\cl}^3M_{\cl}$ (reflecting the Pauli
principle) which they obtained by modeling a cluster of mass $M_{\cl}$
as a square potential well of core radius $r_{\cl}$ filled with a
Fermi-Dirac gas of neutrinos at zero temperature. Subsequently
Tremaine \& Gunn \cite{Tremaine:we} noted that this provides a {\em
lower} bound on the neutrino mass. Although the microscopic
phase-space density (\ref{fnu}) is conserved for collisionless
particles, the `coarse-grained' phase-space density in bound objects
can decrease below its maximum value of $g_\nu/2(2\pi)^3$ during
structure formation. Modeling the bound system as an isothermal sphere
with velocity dispersion $\sigma$ and core radius
$r_{\cl}^2=9\sigma^2/4\pi\,G_{\N}\rho(r_{\cl})$ then gives
\begin{equation}
\label{tg}
 m_\nu > 120 \eV \left(\frac{\sigma}{100\,\km\,\sec^{-1}}\right)^{-1/4} 
         \left(\frac{r_{\cl}}{\kpc}\right)^{-1/2}.
\end{equation}
This is indeed consistent with the cosmological upper bound
(\ref{mnu}) down to the scale of galaxies, however there is a conflict
for smaller objects, viz. dwarf galaxies which require a {\em minimum}
mass of $\sim100$~eV \cite{Spergel:wr,l89}. In fact the central phase
space density of observed dark matter cores in these structures
decreases rapidly with increasing core radius, rather than being
constant as would be expected for neutrinos
\cite{Burkert:1997fz,ss97,Sellwood:2000nh}. Moreover since neutrinos
would cluster more efficiently in larger potential wells, there should
be a trend of increasing mass-to-light ratio with scale. This was
indeed claimed to be the case initially \cite{Schramm:1980xv} but
later it was recognised that the actual increase is far less than
expected \cite{Blumenthal:1984bp}. Thus massive neutrinos are now
disfavoured as the constituent of the `missing mass' in galaxies and
clusters.

\section{The rise and fall of `hot dark matter'}

Nevertheless such cosmological arguments became of particular interest
in the 1980's after the ITEP tritium $\beta$-decay experiment claimed
a $\sim30$~eV mass for the electron neutrino. The attention of
cosmologists turned to how the large-scale structure (LSS) of
galaxies, clusters and superclusters would have formed if the universe
is indeed dominated by such massive neutrinos. The basic picture
\cite{Padmanabhan:1993,Peebles:xt} is that structure grows through
gravitational instability from primordial density perturbations; these
perturbations were first detected by the {\sl COsmic Background
Explorer} (COBE) via the temperature fluctuations they induce in the
CMB \cite{Smoot:1992td}. On small scales ($\lessim10\Mpc$) structure
formation is complicated by non-linear gravitational clustering as
well as non-gravitational (gas dynamic) processes but on large scales
gravitational dynamics is linear and provides a robust probe of the
nature of the dark matter.

Density perturbations in a medium composed of relativistic
collisionless particles are subject to a form of Landau damping
(viz. phase-mixing through free streaming of particles from high to
low density regions) which effectively erases perturbations on scales
smaller than the free-streaming length $\sim41\Mpc(m_\nu/30\eV)^{-1}$
\cite{Doroshkevich:1980,Bond:ha,Doroshkevich:tq}. This is essentially
the (comoving) distance traversed by a neutrino from the Big Bang
until it becomes non-relativistic, and corresponds to the scale of
superclusters of galaxies. Thus huge neutrino condensations
(generically in the shape of `pancakes'), containing a mass
$\sim3\times10^{15}M_\odot(m_\nu/30\eV)^{-2}$, would have begun
growing at a redshift $z_{\eq}\sim7\times10^3(m_\nu/30\eV)$ when the
universe became matter-dominated and gravitational instability set
in. This is well before the epoch of (re)combination at
$z_{\rec}\sim10^3$ so the baryons were still closely coupled to the
photons, while the neutrinos were mildly relativistic ($v/c\sim0.1$)
hence `hot'. After the universe became neutral, baryonic matter would
have accreted into these potential wells, forming a thin layer of gas
in the central plane of the pancakes. Thus superclusters would be the
first objects to condense out of the Hubble flow in a `hot dark
matter' (HDM) cosmogony, and smaller structures such as galaxies would
form only later through the fragmentation of the pancakes.

The gross features of such a `top-down' model for structure formation
are compatible with several observed features of LSS, in particular
the distinctive `voids' and `filaments' seen in large galaxy
surveys. It was also noted that since primordial density perturbations
can begin growing earlier than in an purely baryonic universe, their
initial amplitude must have been smaller, consistent with extant
limits on the isotropy of the microwave background. Detailed studies
\cite{Peebles:ib,Klypin:1983,White:yj,Bond:hb} found however that
galaxies form too late through the breakup of the pancakes, at a
redshift $z\lessim1$, counter to observations of galaxies and quasars
at $z>4$. (Another way of saying this is that galaxies should have
formed {\em last} in an HDM universe, whereas our Galaxy is in fact
dynamically much older than the local group \cite{Peebles:1984}.)
There are other difficulties such as too high `peculiar'
(i.e. non-Hubble) velocities \cite{Kaiser:1983}, excessive X-ray
emission from baryons which accrete onto neutrino clusters
\cite{White:84}, and too large voids \cite{Zeng:1991} (although
detailed simulations \cite{Braun:1987uc,Centrella:1988,Cen:1992}
showed later that some of these problems had perhaps been
exaggerated).

Therefore cosmologists abandoned HDM and turned, with considerably
more success, to cold dark matter (CDM) , i.e. particles which were
non-relativistic at the epoch of matter-domination
\cite{Peebles:1982ff,Bond:fp}. Detailed studies of CDM universes gave
excellent agreement with observations of galaxy clustering and a
`standard CDM' model for large-scale structure formation was
established, viz. a critical density CDM dominated universe with an
initially scale-invariant spectrum of density perturbations
\cite{Davis:rj,cdmrev}. Moreover particle physicists provided
plausible candidate particles, notably the neutralino in
supersymmetric models with conserved $R$-parity which naturally has a
relic abundance of order the critical density \cite{Jungman:1995df}.

\section{COBE and the advent of `mixed dark matter'}

Nevertheless neutrinos were resuscitated some years later as a
possible (sub-dominant) component of the dark matter when the CDM
cosmogony itself ran into problems. To appreciate the background to
this it is necessary to recapitulate the essential ingredients of a
model for cosmic structure formation. A key assumption made
concerns the nature of the primordial density perturbations which grow
through gravitational instability in the dark matter. Cosmologists
usually assume such fluctuations to have a power spectrum of the
{\em scale-free} form:
\begin{equation}
 P (k) = \langle|\delta_k|^2\rangle = A k^n ,
\end{equation}
with $n=1$ corresponding to the scale-invariant `Harrison-Zel'dovich'
spectrum. Here
$\delta_k\equiv\int\case{\delta\rho(\vec{x})}{\bar{\rho}}{\rm
e}^{-i\vec{k}\cdot\vec{x}}{\df}^3x$ is the Fourier transform of
spatial fluctuations in the density field (of wavelength
$\lambda=2\pi/k$). Moreover the perturbations are assumed to be
gaussian (i.e. different phases in the plane-wave expansion are
uncorrelated) and to be `adiabatic' (i.e. matter and radiation
fluctuate together). Concurrent with the above studies concerning the
nature of the dark matter, powerful support for this conjecture was
provided by the development of the `inflationary universe' model
\cite{Kolb:vq,Linde:nc}. Here the perturbations arise from quantum
fluctuations of a scalar field $\phi$, the vacuum energy of which
drives a period of accelerated expansion in the early universe. The
classical density perturbation thus generated has a spectrum
determined by the `inflaton' potential $V(\phi)$, with a power-law
index which is dependent on $k$ \cite{Liddle:cg}:
\begin{equation}
\label{nk}
 n (k) = 1 - 3 M^2 \left(\frac{V'}{V}\right)^2_\star
           + 2 M^2 \left(\frac{V''}{V}\right)_\star
\end{equation}
where $M\equiv(8\pi\,G_{\N})^{-1/2}\simeq2.4\times10^{18}\GeV$ is the
Planck mass and $\star$ denotes that this is to be evaluated when a
mode of wavenumber $k$ crosses the Hubble radius $H^{-1}$. For a
sufficiently `flat potential' (as is necessary to drive enough e-folds
of inflation to solve the problems of the standard cosmology), the
spectrum indeed has $n\simeq1$, with corrections $\propto\ln(k)$.

Gravitational instability sets in only when the universe becomes
matter-dominated and this modifies the spectrum on length scales
smaller than the Hubble radius at this epoch, viz. for
$k>k_{\eq}^{-1}\simeq\,80h^{-1}\Mpc$. Thus the characteristics of the
dark matter can be encoded into a `transfer function' $T(k)$ which
modulates the primordial spectrum; for HDM this is an exponentially
falling function while for CDM it is a more gradual power-law. Now the
power spectrum inferred from observations may be compared with
theoretical models, but another problem arises concerning how we are
to normalize the amplitude of the primordial density
perturbations,\footnote{This is determined by the inflaton potential
$V(\phi)$ but there is, as yet, no `standard model' of inflation
\cite{Liddle:cg}.}  particularly since these are in the dark matter
and may differ significantly (i.e. be `biased') from the observable
fluctuations in the density of visible galaxies. Fortunately, the
primordial perturbations have another unique observational signature,
viz.  they induce temperature fluctuations in the CMB through the
`Sachs-Wolfe effect' (gravitational red/blue shifts) on large angular
scales $\gtrsim2^{0}$, corresponding to spatial scales larger than the
Hubble radius on the last scattering surface \cite{Sachs:er}. It was
the COBE measurement of these fluctuations a decade ago that initiated
the modern era of cosmological structure formation studies.

The quadrupole anisotropy in the CMB measured by COBE
\cite{Smoot:1992td} allows a determination of the fluctuation
amplitude at the scale, $H_0^{-1}\simeq\,3000\,h^{-1}\Mpc$,
corresponding to the present Hubble radius. With this normalization it
became clear that a $\Omega_\nu\simeq1$ HDM universe indeed had too
little power on small-scales for adequate galaxy
formation.\footnote{To save HDM would require new sources of
small-scale fluctuations, e.g. relic topological defects
\cite{Brandenberger:1987er,hdmdef,Gratsias:uc}, or {\em isocurvature}
primordial perturbations \cite{hdmiso,Sugiyama:1988qc} --- to date
however there is no evidence for either.} However it also became
apparent that the `standard CDM model' when normalized to COBE had too
much power on small-scales (see Fig.~1). It was thus a logical step to
invoke a suitable mixture of CDM and HDM to try and match the
theoretical power spectrum to the data on galaxy clustering and
motions \cite{Wright:tf,Davis:ui,Taylor:zh}.

\begin{figure}[tbh]
\label{wss_lss}
\centering
\includegraphics[width=0.8\textwidth]{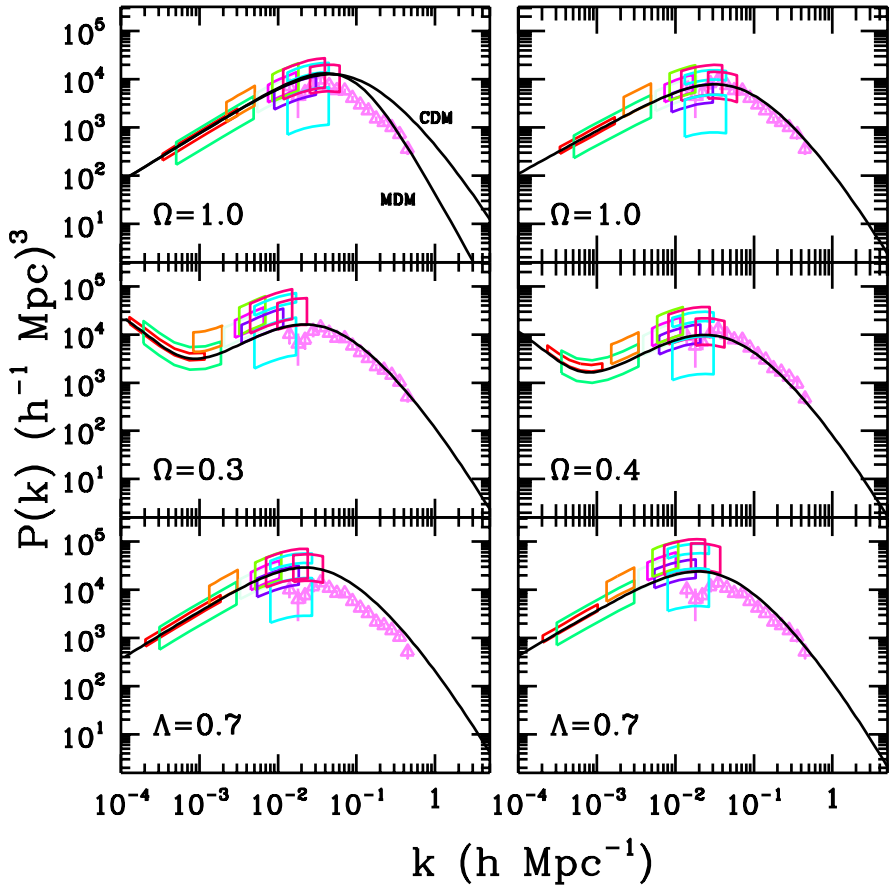}
\caption{The matter power spectrum as inferred from LSS and CMB data
is compared with the CDM cosmogony and several variants in this figure
reproduced from Scott, White and Silk \protect\cite{Scott:1995uj}. We
see top left that the excess small-scale power in the COBE-normalized
`standard' CDM model with scale-invariant adiabatic fluctuations
($n=1, \Omega_{\B}=0.03, h=0.5$), is reduced in the MDM model which
has a substantial neutrino component ($\Omega_\nu=0.3$). This can also
be achieved by `tilting' the spectrum to $n=0.9$ as shown top right
($\Omega_{\B}=0.1, h=0.45$). The middle panels show the expectations
in open universes, while the bottom panels correspond to a flat
universe with a cosmological constant (the bottom right figure assumes
a significant gravitational wave contribution to the COBE signal).}
\end{figure}

In fact the possibility that the dark matter may have both a hot and a
cold component had been discussed several years earlier by Shafi \&
Stecker, motivated by theoretical considerations of SUSY GUTs
\cite{cphdm}, and the possibilities for solving the problems of the
standard CDM model had been noted \cite{Schaefer:ua}. In the post-COBE
era, a number of detailed studies of mixed dark matter (MDM) universes
were performed and a neutrino fraction of about $20\%$ was found to
give the best match with observations
\cite{cphdm2,Klypin:1992sf,Jing:xf,Ma:xs,Pogosian:1994ns,Liddle:1995ay}.
The implied neutrino mass was $\sim5\eV$, presumably that of the
$\nu_\tau$ given the usual hierarchy implied by the see-saw mechanism
\cite{Shafi:1996ha}. The suppression of small-scale power delays the
epoch of galaxy formation, so constraints on the HDM component can
also be obtained from the abundance of high redshift objects such as
QSOs and Ly-$\alpha$ systems; this gave an {\em upper} bound of 4.7~eV
on the neutrino mass \cite{Ma:1994ub}. More baroque schemes in which
two neutrinos have comparable masses
($m_{\nu_\mu}\sim\,m_{\nu_\tau}\sim2.5\eV$) were also constructed
\cite{Primack:1994pe} seeking to reconcile the LSND report of neutrino
oscillations \cite{Athanassopoulos:1995iw} with the atmospheric
neutrino anomaly which had just been reported by Kamiokande
\cite{Fukuda:1994mc}. This was an exciting time for neutrino cosmology
as both laboratory data and astronomical observations supported the
possibility that a substantial fraction of the cosmological mass
density is in the form of massive neutrinos.

However another way to reconcile a CDM universe with the small-scale
observations is to relax the underlying assumption that the primordial
spectrum is strictly {\em scale-invariant}. As shown in Fig.1, a
`tilted' spectrum with $P(k)\propto\,k^{0.9}$ also gives a good fit to
the data \cite{White:1995vk}. At first sight this might strike one as
simply introducing an additional parameter (although this is arguably
no worse than introducing an additional form of dark matter). However
one should really ask why the spectrum should be assumed to have a
power-law index $n=1$ in the first place. As indicated in
Eq.(\ref{nk}), $n$ in fact varies slowly with $k$ and is determined by
the slope and curvature of the scalar potential at the epoch when the
fluctuation at a specified value of $k$ crosses the Hubble radius. The
corresponding number of e-folds before the end of inflation is just
$N_\star(k)\simeq51+\ln\left(\case{k^{-1}}{3000h^{-1}\Mpc}\right)$,
for typical choices of the inflationary scale, reheat temperature
etc. We see that fluctuations on the scales ($\sim1-3000\Mpc$) probed
by LSS and CMB observations are generated just $40-50$ e-folds before
the end of inflation. It is quite natural to expect the inflaton
potential to begin curving significantly as the end of inflation is
approached, especially in `new inflation' (small field) models. There
are certainly attractive models of inflation in which the spectrum is
significantly tilted in this region, in particular an inflationary
model based on $N=1$ supergravity \cite{Ross:1995dq} naturally gives
$n(k)\simeq(N_\star-2)/(N_\star+2)\sim0.9$ at these scales. With such
a {\em scale-dependent} tilt for the primordial spctrum, the LSS data
can be fitted reasonably well, with no need for any HDM component
\cite{Adams:1996yd}. (Of course in the absence of a `standard model'
of inflation, it might be argued that the inflationary spectrum may
instead have $n>1$ thus allowing a {\em larger} HDM component.)

As Fig.1 shows, yet another way of suppressing small-scale power in
the CDM cosmogony is to decrease the matter content of the universe,
since this postpones the epoch of matter-radiation equality and thus
shifts the peak of power spectrum to larger scales. Furthermore the
spatial geometry can be maintained flat if there is a cosmological
constant with $\Omega_\Lambda=1-\Omega_{\m}\sim0.7$.  Evidence for
just such a cosmology ($\Lambda$CDM) has come subsequently from
observations of the Hubble diagram of Type~Ia supernovae which suggest
that the expansion is in fact {\em accelerating}
\cite{Riess:1998cb,Perlmutter:1998np}, coupled with the observation
that $\Omega_{\m}$ does not exceed $\sim0.3$ even on the largest
scales probed \cite{Bahcall:1999xn}.  Assuming such a $\Lambda$CDM
cosmology and a scale-invariant power spectrum gives a limit of
$f_\nu\equiv\Omega_\nu/\Omega_{\m}<0.13$ ($\95cl$) using the power
spectrum of galaxy clustering determined from the {\sl Two degree
Field galaxy redshift survey} (2dFGRS) as shown in Fig.~2. This
corresponds to an upper bound of $\sum m_\nu < 1.8~\eV (\95cl)$ on the
sum of neutrino masses, adopting `concordance' values of $\Omega_{\m}$
and $h$ \cite{Elgaroy:2002bi}. If the spectral index is allowed to
vary in the range $n=1\pm0.1$, then this bound is relaxed to 2.1~eV.

\begin{figure}[tbh]
\label{2df}
\centering \includegraphics[width=0.5\textwidth]{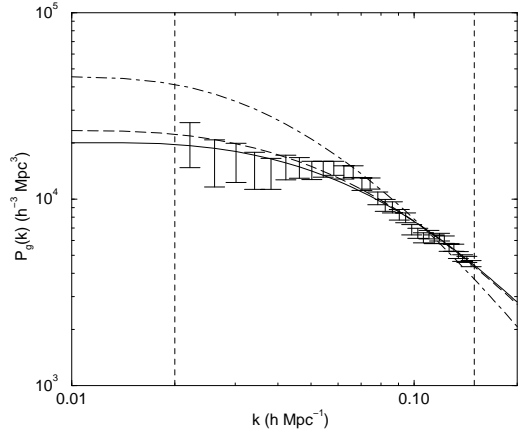}
\caption{The 2dFGRS power spectrum of galaxy clustering is compared
with the expectations for no neutrino dark matter (solid line),
$\Omega_\nu=0.01$ (dashed line) and $\Omega_\nu=0.05$ (dot-dashed
line). A scale-invariant spectrum is assumed and the cosmological
parameters adopted are $\Omega_{\m}=0.3$, $\Omega_\Lambda=0.7$,
$h=0.7$, $\Omega_{\B}h^2=0.02$ (from Elgaroy \etal
\protect\cite{Elgaroy:2002bi}).}
\end{figure}

\section{CMB anisotropy and limits on neutrino dark matter} 

It is clear that there are uncertainties in the determination of the
HDM component from LSS data alone. Fortunately it is possible to
reduce these substantially by examining an independent probe of the
primordial power spectrum, viz.  temperature fluctuations in the CMB.

In general a skymap of the CMB temperature can be decomposed into
spherical harmonics
\begin{equation}
 T (\theta, \phi) = \sum_{l=0}^{\infty} \sum_{m=-l}^{l}
                     a_{l}^{m} Y_{l}^{m} (\theta, \phi) ,
\end{equation} 
where the $l^{\rm th}$ multipole corresponds to an angle
$\theta^{\circ}\sim200/l$ and probes spatial scales around
$k^{-1}\sim6000h^{-1}l^{-1}\Mpc$. In inflationary theories, the
fluctuations are gaussian so the co-efficients $a_{l}^{m}$ are
independent stochastic variables with zero mean and variance
$C_{l}=\langle|a_{l}^{m}|^2\rangle$; each $C_l$ has a $\chi^2$
distribution with $(2l+1)$ degrees of freedom
\cite{Peebles:1982ff,Abbott:1983mk}. For an assumed set of
cosmological parameters and given the primordial density perturbation
spectrum, the $C_{l}$'s can be determined by solution of the
Einstein-Boltzmann equations which describe how the different
components (photons, ions, electrons, neutrinos \ldots)
evolve \cite{Hu:1995fq,Bertschinger:1995er,Seljak:1996is}. Thus
theoretical estimates of the power at each multipole can be compared
with observations. The low multipoles (large spatial scales) are
sensitive to the primordial spectrum alone,\footnote{In principle,
primordial gravitational waves can also make a contribution here but
this is expected to be negligible in `small-field' inflationary
models \cite{Ross:1995dq,Lyth:1996we}.} but the measurements in the
region are particularly uncertain, both because of uncertainties in
the foreground substraction and also because there are fewer
independent measurements for low multipoles (`cosmic variance'). For
example, by measuring the first $\sim$20 multipoles COBE could only
fix $n=1.2\pm0.3$ \cite{Smoot:1992td} so could not discriminate
between a scale-invariant and a mildly tilted spectrum. However
subsequent ground-based experiments with angular resolution far
superior to COBE's have now measured the power at higher multipoles
\cite{Hu:2001bc}. The dominant features in the power spectrum here are
the `acoustic peaks', the most prominent being at $l\sim200$, arising
from oscillations of the coupled plasma-photon fluids at last
scattering \cite{Hu:1996qs}. The position of the first peak is a
measure of the horizon length at the epoch of (re)combination of the
primordial plasma and thus provides a measure of space curvature ---
observations by the BOOMERanG \cite{deBernardis:2000gy} and MAXIMA
\cite{Balbi:2000tg} experiments were the first to show that this is in
fact close to zero. Taken together with the earlier recognition that
CDM alone cannot make up the critical density, this has led to the
widespread adoption of the $\Lambda$CDM model in which $\Omega_{\m}\sim0.3$,
$\Omega_\Lambda\sim0.7$ \cite{Bahcall:1999xn,Sahni:1999gb}.

\begin{figure}[tbh]
\label{mdm_cl}
\centering
\includegraphics[width=0.5\textwidth,height=0.25\textheight]{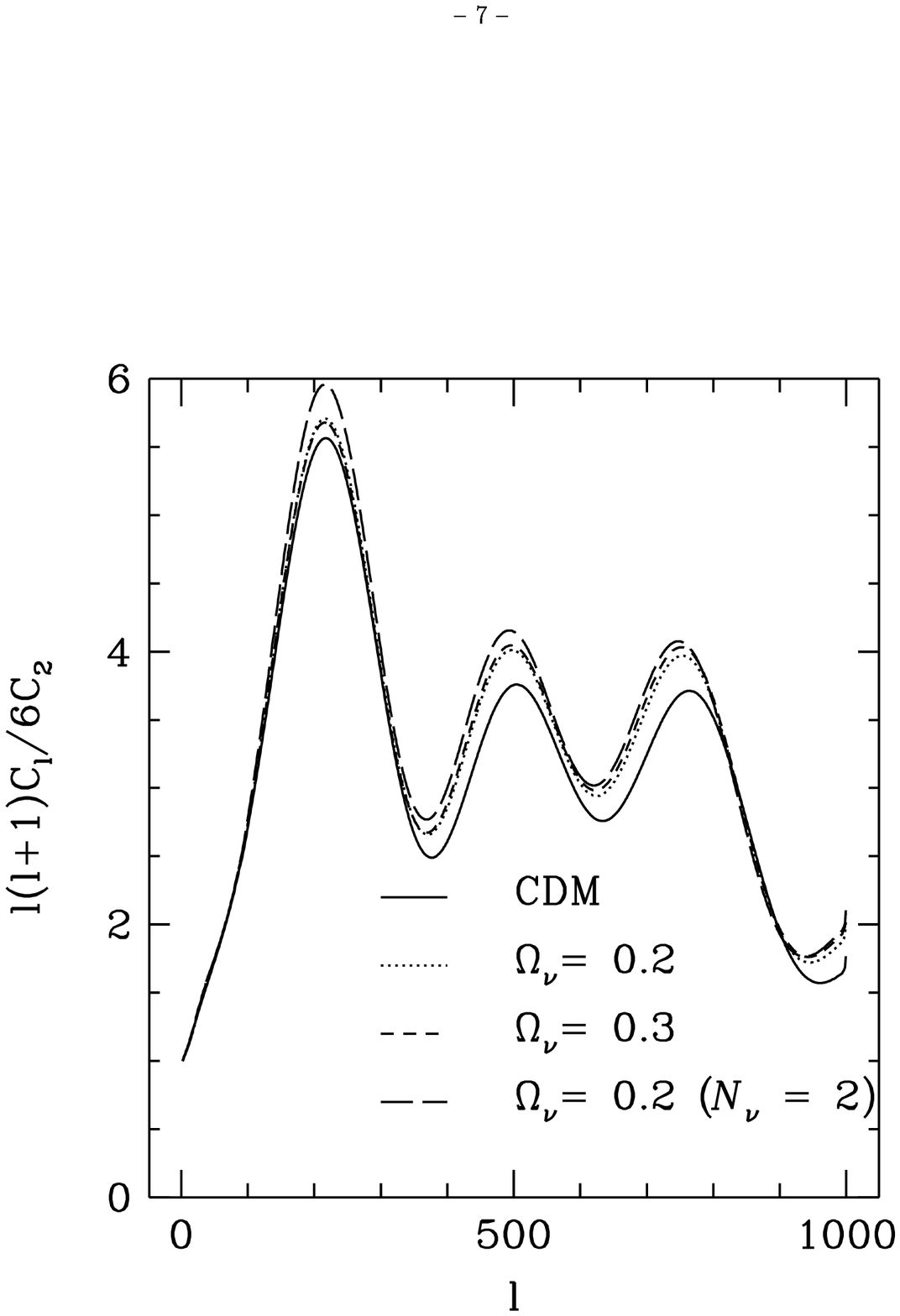}
\caption{The angular power spectrum of CMB anisotropy (assuming a
scale-invariant spectrum) in CDM and MDM universes, as calculated by
Dodelson, Gates and Stebbins \protect\cite{Dodelson:1995es}.}
\end{figure}

As seen in Fig.~3, the expectations for CMB anisotropy in a MDM
universe do not differ significantly from a CDM universe having the
same initial perturbation spectrum. Thus it is clear that by combining
CMB and LSS data, it would be possible to determine whether the
suppression of small-scale power is {\em intrinsic} to the primordial
spectrum of inflationary fluctuations, or is induced by a HDM
component. However there are additional uncertainties in the input
values of the other cosmological parameters, in particular the values
of the Hubble parameter and of the baryon density which is usually
inferred from considerations of primordial nucleosynthesis. In
analysing the observational data, various `priors' are often assumed
for these quantities on the basis of other observations. For example a
detailed likelihood analysis \cite{Wang:2001gy} yielded the neutrino
density fraction $f_\nu$ to be at most 0.08 from CMB observations
alone, decreasing to 0.06 if LSS data from the PSCz survey is added,
and further to 0.04 if the value of $h=0.72\pm0.08$
\cite{Freedman:2000cf} is adopted. Another analysis
\cite{Hannestad:2002xv} used the more precise 2dFGRS data
\cite{Elgaroy:2002bi}, together with additional constraints on the
matter density from the SNIa data ($\Omega_{\m}=0.28\pm0.14$ assuming
a flat universe) \cite{Riess:1998cb,Perlmutter:1998np}, and on the
baryon density from primordial nucleosynthesis
($\Omega_{\B}h^2=0.02\pm0.002~(\95cl)$ \cite{Burles:2000zk}); using
the analytic result for the suppression of the power spectrum, $\Delta
P/P\simeq-8f_\nu$ \cite{Hu:1997mj}, a bound of $\sum
m_\nu<2.5-3~{\eV}$ was obtained.

\begin{figure}[tbh]
\label{mnu_lim}
\centering
\includegraphics[width=0.48\textwidth]{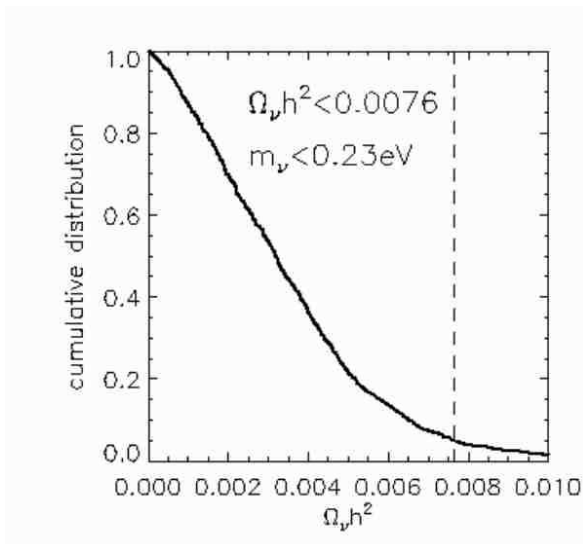}
\caption{The marginalized cumulative probability of $\Omega_{\nu}h^2$
is shown, based on a fit to the WMAP data on CMB anisotropies,
together with the 2DFGRS data on of galaxy clustering (from Spergel
\etal \protect\cite{Spergel:2003cb}, courtesey of WMAP Science Team).}
\end{figure}

It had been estimated \cite{Hu:1997mj} that the CMB data expected from
the {\sl Wilkinson Microwave Anisotropy Probe} (WMAP)
\cite{Bennett:2003ba}, combined with LSS data from the Sloane Digital
Sky Survey (SDSS) \cite{Dodelson:2001ux}, will provide sensitivity to
neutrino mass at the eV level. Indeed the recently announced first
results from WMAP, combined with 2dFGRS, have already set the bound
$\Omega_{\nu}h^2<0.0076$ corresponding to $\sum m_\nu<0.7~{\eV}
(\95cl)$, as shown in Fig.~4 \cite{Spergel:2003cb}. This severely
restricts \cite{Pierce:2003uh}, but does not altogether rule out
\cite{Giunti:2003cf}, a fourth (sterile) neutrino with a mass of
${\cal O}(1)$~eV as suggested by the LSND experiment
\cite{Athanassopoulos:1995iw}. Combined with the new KamLAND data,
this also restricts the accessible range for the observation of
neutrinoless $\beta\beta$ decay \cite{Bhattacharyya:2003gi}, which has
been claimed to have been seen already implying an effective Majorana
mass of $|\langle\,m_\nu\rangle|= 0.39^{+0.45}_{-0.34}\eV~(\95cl)$
\cite{Klapdor-Kleingrothaus:md}.

\section{The nucleosynthesis limit on $N_\nu$}

Hoyle \& Taylor \cite{ht64} as well as Peebles \cite{p66} had
emphasized many years ago that new types of neutrinos (beyond the
$\nu_e$ and $\nu_\mu$ then known) would boost the relativistic energy
density hence the expansion rate (\ref{H}) during Big Bang
nucleosynthesis (BBN), thus increasing the yield of $\he4$. Shvartsman
\cite{Shvartsman:1969mm} noted that new superweakly interacting
particles would have a similar effect. Subsequently this argument was
refined quantitatively by Steigman, Schramm \& Gunn
\cite{Steigman:kc}. In the pre-LEP era when the laboratory bound on
the number of neutrino species was not very restrictive
\cite{Denegri:1989if}, the BBN constraint was used to argue that at
most one new family was allowed \cite{Yang:gn,Steigman:1986nh}, albeit
with considerable uncertainties \cite{Ellis:1985fp}. Although LEP now
finds $N_\nu=2.994\pm0.012$ \cite{Hagiwara:fs}, the cosmological bound
is still important since it is sensitive to {\em any} new light
particle \cite{Steigman:xp}, not just $SU(2)_{\L}$ doublet neutrinos,
so is a particularly valuable probe of new physics, e.g. neutrinos
coupled to new gauge bosons expected in string models
\cite{Ellis:1985fp,Barger:2003zh}.

BBN limits on neutrinos come mainly from the observational bounds on
the primordial $\he4$ abundance, termed $Y_\pr$ by astronomers. This
is proportional to the neutron fraction which `freezes out' at
$n/p\simeq\exp[{-(m_n-m_p)/T_{\fr}}]$ when the weak interaction rate
($\propto\,G_{\F}^2T^5$) falls behind the Hubble expansion rate
(\ref{H}), at $T_{\fr}\sim(g_*G_{\N}/G_{\F}^4)^{1/6}\approx1$~MeV. The
presence of additional neutrino flavors (or of any other relativistic
species) at this time increases $g_*$, hence the expansion rate $H$,
leading to a larger value of $T_\fr$, $n/p$, and thus $Y_\pr$. In the
Standard Model, the number of relativistic particle species at 1~MeV
is
\begin{equation}
 g_* = 5.5 + \case{7}{4} N_\nu,
\label{g_*}
\end{equation}
where 5.5 accounts for photons and $e^{\pm}$, and $N_\nu$ is the
number of (massless) neutrino flavors. (The energy density of new
light fermions $i$ is equivalent to an effective number
$\Delta\,N_{\nu}=\sum_{i}(g_i/2)(T_i/T_\nu)^4$ of additional doublet
neutrinos, where $T_i/T_\nu$ follows from considerations of their
(earlier) decoupling.) In Fig.~5, we show the expected abundance of
$\he4$ taking $N_\nu=3$, as a function of the density of baryons
normalized to the blackbody photon density:
\begin{equation}
\eta \equiv \frac{n_{\B}}{n_\gamma} \simeq 2.728\times10^{-8}\Omega_{\B}h^2, 
\label{eta}
\end{equation}
with $\eta_{10}\equiv\eta/10^{-10}$. The computed $\he4$ abundance
scales linearly with $\Delta\,N_\nu$ but it also increases
logarithmically with $\eta$ \cite{Bernstein:ad,Sarkar:1995dd}:
\begin{equation}
\Delta Y \simeq 0.012 \Delta N_\nu 
              + 0.01 \ln \left(\frac{\eta_{10}}{5}\right).
\label{Y}
\end{equation}
Thus to obtain a bound on $N_\nu$ requires an upper limit on $Y_{\pr}$
{\em and} a lower limit on $\eta$. The latter is poorly determined
from direct observations of luminous matter \cite{Fukugita:1997bi} so
must be derived from the abundances of the other synthesized light
elements, $\h2$, $\hel3$ and $\li7$, which are power-law functions of
$\eta$ as seen in Fig.~5. The complication is that these abundances
are substantially altered in a non-trivial manner during the chemical
evolution of the galaxy, unlike $Y_{\pr}(\he4)$ which just increases
by a few percent due to stellar production. (This can be tagged via
the correlated abundance of oxygen and nitrogen which are made {\em
only} in stars.)  Even so, chemical evolution arguments have been used
to limit the primordial abundances of $\h2$ and $\hel3$ and thus
derive increasingly severe bounds on $N_\nu$
\cite{Olive:1989xf,Walker:ap,Copi:ev}, culminating in one {\em below}
3 \cite{Hata:1995tt}! However a more conservative view
\cite{Kernan:1996yz} is that there is no `crisis' with BBN if we
recognize that chemical evolution arguments are unreliable and
consider only {\em direct} measurements of light element abundances.

\begin{figure}[tbh]
\label{bbn}
\centering
\includegraphics[width=0.68\textwidth]{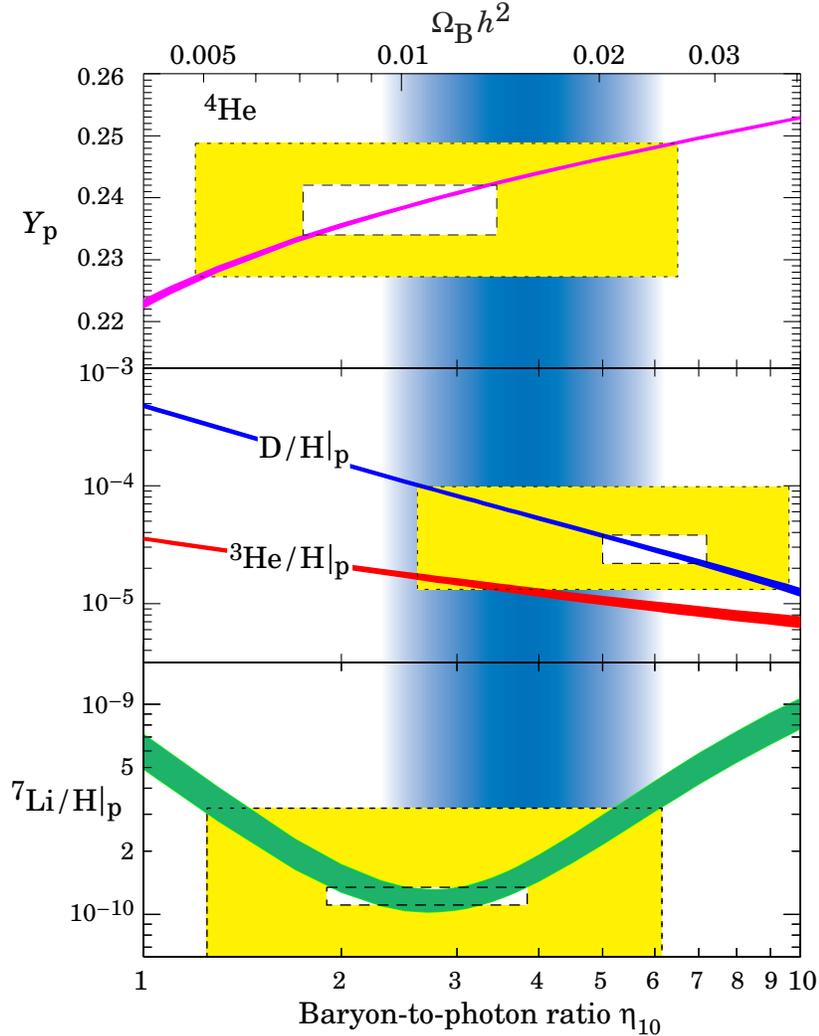}
\caption{The primordial abundances of $\he4$, $\h2$, $\hel3$ and $\li7$
as predicted by the standard BBN model compared to observations ---
smaller boxes: $2\sigma$ statistical errors; larger boxes: $\pm
2\sigma$ statistical and systematic errors added in quadrature (from
Fields and Sarkar \protect\cite{Hagiwara:fs}).}
\end{figure}

Fig.~5 shows the current status of such measurements, as reviewed in
more detail elsewhere \cite{Hagiwara:fs,Sarkar:2002er}. We observe $\he4$
in clouds of ionized hydrogen (H~II regions), the most metal-poor of
which are in dwarf blue compact galaxies (BCGs). There is now a large
body of data on $\he4$ and C, N, O in these systems which confirm that
the small stellar contribution to helium is positively correlated with
`metal' production; extrapolating to zero metallicity gives the
primordial $\he4$ abundance \cite{fo}:
\begin{equation}
 Y_\pr = 0.238 \pm 0.002 \pm 0.005 \ . 
\label{YpObs} 
\end{equation}
Here and subsequently, the first error is statistical, and the second is
an estimate of the systematic uncertainty. The latter clearly
dominates, and is based on the scatter in different analyses of the
physical properties of the H~II regions \cite{os,Sauer:2001vh}. Other
extrapolations to zero metallicity give $Y_\pr=0.244\pm0.002$
\cite{izo}, and $Y_\pr=0.235\pm0.003$ \cite{ppr}, while the average in
the 5 most metal-poor objects is $Y_\pr=0.238\pm0.003$ \cite{ppl}. The
value in Eq.(\ref{YpObs}), shown in Fig.~5, is consistent with all
these determinations.

The systems best suited for Li observations are hot, metal-poor stars
belonging to the halo population (Pop~II) of our Galaxy. Observations
have long shown that Li does not vary significantly in such stars
having metallicities less than 1/30 of Solar --- the `Spite
plateau'.  Recent precision data suggest a small but significant
correlation between Li and Fe \cite{rnb} which can be understood as
the result of Li production from cosmic rays
\cite{sfosw}. Extrapolating to zero metallicity one arrives at a
primordial value \cite{rbofn}
\begin{equation}
{\rm Li/H}|_\pr =
 (1.23 \pm 0.06 {}^{+0.68}_{-0.32} {}^{+0.56}) \times 10^{-10} \ .
\label{Lip}
\end{equation}
The last error is our estimate of the maximum upward correction
necessary to allow for possible destruction of Li in Pop~II stars,
due to e.g. mixing of the outer layers with the hotter interior
\cite{vau,ddk}. Such processes can be constrained by the absence of
significant scatter in the Li-Fe correlation plot \cite{rnb}, and
through observations of the even more fragile isotope $^6$Li
\cite{sfosw}.

In recent years, high-resolution spectra have revealed the presence of
D in quasar absorption systems (QAS) at high-redshift, via its
isotope-shifted Lyman-$\alpha$ absorption. It is believed that there
are no astrophysical sources of deuterium, so any measurement of D/H
provides a lower limit to the primordial abundance and thus an upper
limit on $\eta$; for example, the local interstellar value of
$\h2/\hy=(1.5\pm0.1)\times10^{-5}$ \cite{lin} requires
$\eta_{10}\le9$. Early reports of $\h2/\hy>10^{-4}$ towards 2 quasars
(Q0014+813 \cite{schr} and PG1718+4807 \cite{webb}) have been
undermined by later analyses \cite{bkt,Kirkman:2001nk}. Three high
quality observations yield $\h2/\hy=(3.2\pm0.3)\times10^{-5}$
(PKS1937-1009), $(4.0\pm0.7)\times10^{-5}$ (Q1009+2956), and
$(2.5\pm0.2)\times10^{-5}$ (HS0105+1619); their average value
\begin{equation}
\h2/\hy = (3.0\pm0.4)\times10^{-5}
\label{D}
\end{equation}
has been widely promoted as {\em the} primordial abundance
\cite{ome}. Recently, the same group \cite{Kirkman:2003uv} have
provided another measurement:
$\h2/\hy=(2.42^{+0.35}_{-0.25})\times10^{-5}$ (HS 243+3057). However
the observed dispersion in the measurements suggests either that
systematic uncertainties have been underestimated, or that there is
intrinsic dispersion in the D abundance in QAS. Other values have been
reported in different (damped Lyman-$\alpha$) systems which have a
higher column density of neutral H,
viz. $\h2/\hy=(2.24\pm0.67)\times10^{-5}$ (Q0347-3819) \cite{dod} and
$\h2/\hy=(1.65\pm0.35)\times10^{-5}$ (Q2206-199) \cite{pb}. Moreover,
allowing for a more complex velocity structure than assumed in these
analyses raises the inferred abundance by upto $\sim50\%$
\cite{lev}. Even the ISM value of D/H now shows unexpected scatter of
a factor of 2 \cite{son}. All this may indicate significant processing
of the $\h2$ abundance even at high redshift. Given these
uncertainties, it is prudent to bound the primordial abundance with an
upper limit set by the non-detection of D absorption in a
high-redshift system (Q0130-4021) \cite{ktblo}, and the lower limit
set by the local interstellar value \cite{lin}, both at $2\sigma$:
\begin{equation}
1.3 \times 10^{-5} < \h2/\hy|_\pr < 9.7 \times 10^{-5} .
\label{Dp}
\end{equation}

For $\hel3$, the only observations available are in the Solar system and
(high-metallicity) H~II regions in our Galaxy \cite{bbrw}. This makes
inference of the primordial abundance difficult, a problem compounded
by the fact that stellar nucleosynthesis models for $\hel3$ are in
conflict with observations \cite{dst}. Such conflicts can perhaps be
resolved if a large fraction of low mass stars destroy $\hel3$ by
internal mixing driven by stellar rotation, consistent with the
observed $^{12}$C/$^{13}$C ratios \cite{char}. The observed abundance
`plateau' in H~II regions then implies a limit on the primordial
value of $\hel3/\hy<(1.9\pm0.6)\times10^{-5}$ \cite{bbrw}, which is
consistent with the other abundance constraints we discuss.

The overlap in the $\eta$ ranges spanned by the larger boxes in Fig.~5
indicates overall concordance between the various abundance
determinations. Accounting for theoretical uncertainties as well as
the statistical and systematic errors in observations, there is
acceptable agreement among the abundances when \cite{Hagiwara:fs}
\begin{equation}
2.6 \le \eta_{10} \le 6.2 \ .
 \label{etarange}
\end{equation}
However the agreement is far less satisfactory if we use only the
quoted statistical errors in the observations. As seen in Fig.~5,
$\he4$ and $\li7$ are consistent with each other but favor a value of
$\eta$ which is {\em discrepant} by $\sim2\sigma$ from the value
$\eta_{10}=5.9\pm0.4$ indicated by the D abundance (\ref{D}) alone. It
is important to note that it is the latter which essentially provides
the widely-quoted `precision' determination of the baryon density:
$\Omega_{\B}h^2=0.02\pm0.002~(\95cl)$ \cite{Burles:2000zk}. Additional
studies are required to clarify if this discrepancy is real. 

The recent WMAP data on CMB fluctuations implies a baryon abundance of
$\eta_{10}=6.5^{+0.4}_{-0.3}$ if the primordial perturbations are
assumed to have a power-law form \cite{Spergel:2003cb}; the value
decreases by $\sim9\%$ if the spectral index is allowed to vary with
scale, as is now indicated by the observations. Although broadly
consistent with the determinations based on primordial abundances, it
is clear that this exacerbates the tension with the $\he4$ and $\li7$
measurements mentioned above. It has been proposed that the CMB
determination of $\eta_{10}$ can be used as an input into BBN
calculations \cite{Cyburt:2001pq}. However it would be advisable to
await further LSS data from SDSS \cite{Dodelson:2001ux} to pin down
more precisely the primordial fluctuation spectrum, to which such
determinations are sensitive. For example, allowing for a `step' in
the primordial spectrum at $k\sim0.05h$~Mpc$^{-1}$, as was indicated
by data from the APM galaxy survey \cite{Adams:1997de,Barriga:2000nk},
can {\em decrease} the baryon density inferred from the WMAP+2DFGRS
data by upto $\sim40\%$ \cite{Bridle:2003sa}.

Given this situation, it is still neccessary to be conservative in
evaluating bounds on $N_\nu$ from BBN. An analysis based on simple
$\chi^2$ statistics and taking into account the correlated
uncertainties of the elemental yields, gives \cite{Lisi:1999ng}
\begin{equation}
 2 < N _\nu < 4~(\95cl).
\end{equation}
Tighter bounds can of course be obtained under less conservative
assumptions, e.g. adopting the D abundance (\ref{D}) requires
$N_\nu<3.2$ \cite{Burles:1999zt}. If true, such bounds rule out the
possibility of `sterile' neutrinos since these would be brought into
equilibrium through mixing with the active neutrinos in all currently
viable schemes \cite{DiBari:2001ua,Abazajian:2002bj}. The WMAP
determination of the baryon density seems to require $N_\nu$ less than
3 \cite{Hannestad:2003xa} but is consistent with the Standard Model
taking systematic uncertainties in the elemental abundances into
account \cite{Pierce:2003uh,Cyburt:2003fe}. Moreover as emphasized
above, the WMAP determination of $\eta$ will have to ve lowered if the
primordial spectrum is not scale-free and this will considerably ease
the present tension. (The WMAP+2dFGRS data by themselves restrict
$1.8<N_\nu<5.7~(\95cl)$ assuming a flat universe
\cite{Pierpaoli:2003,Crotty:2003th}.)

The above limits hold for the standard BBN model and one can ask to
what extent they can be modified if plausible changes are made to the
model. For example, there may be an initial excess of electron
neutrinos over antineutrinos, parametrised by a `chemical potential'
$\xi_e$ in the distribution function:
$f_\nu\propto[\exp(p/T-\xi_e)+1]^{-1}$. Then $n-p$ equilibrium is
shifted in favour of less neutrons, thus reducing the $\he4$ abundance
by a factor $\approx\exp{(-\xi_e)}$ \cite{Bernstein:ad}. However the
accompanying increase in the relativistic energy density speeds up the
expansion rate and increases the $n/p$ ratio at freeze-out, leading to
more $\he4$, although this effect is smaller. For neutrinos of other
flavours which do not participate in nuclear reactions, only the
latter effect was presumed to operate, allowing the possibility of
balancing a small chemical potential in $\nu_e$ by a much larger
chemical potential in $\nu_{\mu,\tau}$, and thus substantially
enlarging the concordance range of $\eta_{10}$ \cite{degen}. However
the recent recognition from Solar and atmospheric neutrino experiments
that the different flavours are maximally mixed, no longer permits
such a hierarchy of chemical potentials
\cite{Lunardini:2000fy,Dolgov:2002ab,Wong:2002fa,Abazajian:2002qx},
thus ruling out this possible loophole. Consequently the relic
neutrino abundance cannot be significantly different from the value
(\ref{nnu}). A small chemical potential in electron neutrinos of
$\xi_e\lessim0.08$ is however still possible and could be another
explanation for why the $\he4$ abundance seems to be lower than the
value that would be expected on the basis of the measurements of the D
abundance or the CMB anisotropy.

Another possible change to standard BBN is to allow inhomogeneities in
the baryon distribution, created e.g. during the QCD (de)confinement
transition. If the characteristic inhomogeneity scale exceeds the
neutron diffusion scale during BBN, then increasing the average value
of $\eta$ increases the synthesised abundances such that the
observational limits essentially rule out such inhomogeneities.
However fluctuations in $\eta$ on smaller scales will result in
neutrons escaping from the high density regions leading to spatial
variations in the $n/p$ ratio which might allow the upper limit to
$\eta$ to be raised substantially \cite{Applegate:hm}. Recent calculations
show that D and $\he4$ can indeed be matched even when $\eta$ is
raised by a factor of $\sim2$ by suitably tuning the amplitude and
scale of the fluctuations, but this results in unacceptable
overproduction of $\li7$ \cite{inhom2}. A variant on the above
possibility is to allow for regions of antimatter which annihilate
during or even after BBN; however the $\li7$ abundance again restricts
the possibility of raising the limit on $\eta$ substantially
\cite{antimat}. Finally the synthesised abundances can be altered if a
relic massive particle decays during or after BBN generating
electromagnetic and hadronic showers in the radiation-dominated
plasma. Interestingly enough the processed yields of D, $\he4$ and
$\li7$ can then be made to match the observations even for a universe
closed by baryons \cite{desh}, however the production of $^6$Li is
excessive and argues against this possibility. 

In summary, standard BBN appears to be reasonably robust (although
non-standard possibilities cannot be {\em definitively} ruled out) and
consistent with $N_\nu=3$, leaving little room for new
physics. Systematic uncertainties in the elemental abundance
determinations need to be substantially reduced in order to make
further progress.

\section{Conclusions}

The study of neutrinos in cosmology has had a long history. However as
both theory and observations have improved, the fascinating
possibilities that had been raised in early work have gradually been
eliminated. In particular neutrino (hot) dark matter is no longer
required by our present understanding of large-scale
structure. Cosmological bounds on neutrino masses continue to be more
restrictive than laboratory limits and all known species are now
required to be sufficiently light that the rich phenomenology of
unstable neutrinos is now largely of historical interest. Moreover
there is no motivation from considerations of primordial
nucleosynthesis for invoking new sterile neutrinos. Nevertheless such
neutrinos can exist (e.g. in the `bulk'
\cite{Arkani-Hamed:1998vp,Abazajian:2000hw}) and there may yet be
surprises from the Big Bang in store for us.


\begin{thebibliography}{99}

\bibitem{afh53}
R.A.~Alpher, J.W.~Follin \& R.C.~Hermann, 
Phys. Rev. {\bf 92} (1953) 1347.

\bibitem{Hagiwara:fs}
K.~Hagiwara \etal [Particle Data Group Collaboration],
Phys. Rev. D {\bf 66} (2002) 010001.

\bibitem{cm61} 
H.Y.~Chiu \& P.~Morrison, 
Phys. Rev. Lett. {\bf 5} (1960) 573.

\bibitem{z6566} 
Ya.B.~Zel'dovich, 
Adv. Astron. Astrophys. {\bf 3} (1965) 241;
Sov. Phys. Usp. {\bf 9} (1967) 602. 

\bibitem{dg70}
T.~De Graaf, 
Lett. Nuovo\ Cim. {\bf 4} (1970) 638. 

\bibitem{Dicus:bz}
D.A.~Dicus, E.W.~Kolb, A.M.~Gleeson, E.C.~Sudarshan, V.L.~Teplitz \& M.S.~Turner,
Phys. Rev. D {\bf 26} (1982) 2694.

\bibitem{Enqvist:gx}
K.~Enqvist, K.~Kainulainen \& V.~Semikoz,
Nucl. Phys. B {\bf 374} (1992) 392.

\bibitem{Dodelson:1992km}
S.~Dodelson \& M.S.~Turner,
Phys. Rev. D {\bf 46} (1992) 3372.

\bibitem{Hannestad:1995rs}
S.~Hannestad \& J.~Madsen,
Phys. Rev. D {\bf 52} (1995) 1764.

\bibitem{c66}
H.~Chiu, 
Ann. Rev. Nucl. Sci. {\bf 16} (1966) 591.

\bibitem{ps61}
B.~Pontecorvo \& Ya.~Smorodinski, 
Sov. Phys. JETP {\bf 14} (1962) 173.

\bibitem{zs61}
Ya.B.~Zel'dovich \& Ya.~Smorodinski, 
Sov. Phys. JETP\ {\bf 14} (1962) 647.

\bibitem{w62}
S.~Weinberg, 
Phys. Rev. {\bf 128} (1962) 1457.

\bibitem{Gershtein:gg}
S.S.~Gershtein \& Ya.B.~Zeldovich,
JETP Lett. {\bf 4} (1966) 120.

\bibitem{ms72}
G.~Marx \& A.S.~Szalay, 
Proc. Neutrino'72, Balatonf\"urd, Vol.{\bf 1} (1972) p.123;\\
A.S.~Szalay \& G.~Marx, 
Astron. Astrophys. {\bf 49} (1976) 437.

\bibitem{Cowsik:gh}
R.~Cowsik \& J.~McClelland,
Phys. Rev. Lett.  {\bf 29} (1972) 669.

\bibitem{Shapiro:rr}
S.L.~Shapiro, S.A.~Teukolsky \& I.~Wasserman,
Phys. Rev. Lett.  {\bf 45} (1980) 669.

\bibitem{Olive:1981ak}
K.A.~Olive \& M.S.~Turner,
Phys. Rev. D {\bf 25} (1982) 213.

\bibitem{Dolgov:2000ew}
A.D.~Dolgov \& S.H.~Hansen,
Astropart. Phys.  {\bf 16} (2002) 339.

\bibitem{Bernstein:iy}
J.~Bernstein \& G.~Feinberg,
Phys. Lett. B {\bf 101} (1981) 39
[Erratum-ibid. B {\bf 103} (1981) 470].

\bibitem{Kolb:vq}
E.W.~Kolb \& M.S.~Turner,
{\sl The Early Universe} (Addison-Wesley, 1990).

\bibitem{Lee:1977ua}
B.W.~Lee \& S.~Weinberg,
Phys. Rev. Lett.  {\bf 39} (1977) 165.

\bibitem{Vysotsky:pe}
M.I.~Vysotsky, A.D.~Dolgov \& Ya.B.~Zeldovich,
JETP Lett. {\bf 26} (1977) 188.

\bibitem{Sato:1977ye}
K.~Sato \& M.~Kobayashi,
Prog. Theor. Phys.  {\bf 58} (1977) 1775.

\bibitem{Dicus:1977qy}
D.A.~Dicus, E.W.~Kolb \& V.L.~Teplitz,
Astrophys. J.  {\bf 221} (1978) 327.

\bibitem{Enqvist:1988we}
K.~Enqvist, K.~Kainulainen \& J.~Maalampi,
Nucl. Phys. B {\bf 317} (1989) 647.

\bibitem{Hawkins:2002sg}
E.~Hawkins {\it et al.},
arXiv:astro-ph/0212375.

\bibitem{Freedman:2000cf}
W.L.~Freedman {\it et al.} [Hubble Key Project Collaboration],
Astrophys. J.  {\bf 553} (2001) 47.

\bibitem{Weinheimer:2002rs}
C.~Weinheimer,
arXiv:hep-ex/0210050.

\bibitem{Barger:1998kz}
V.D.~Barger, T.J.~Weiler \& K.~Whisnant,
Phys. Lett. B {\bf 442} (1998) 255.

\bibitem{Gonzalez-Garcia:2002dz}
M.C.~Gonzalez-Garcia \& Y.~Nir,
arXiv:hep-ph/0202058.

\bibitem{Osipowicz:2001sq}
A.~Osipowicz {\it et al.}  [KATRIN Collaboration],
arXiv:hep-ex/0109033.

\bibitem{Fukugita:1997bi}
M.~Fukugita, C.J.~Hogan \& P.J.E.~Peebles,
Astrophys.\ J.\  {\bf 503} (1998) 518

\bibitem{cm72b} 
R.~Cowsik \& J.~McClleland, 
Astrophys. J. {\bf 180} (1972) 7.

\bibitem{Tremaine:we}
S.~Tremaine \& J.E.~Gunn,
Phys. Rev. Lett.  {\bf 42} (1979) 407.

\bibitem{Spergel:wr}
D.N.~Spergel, D.H.~Weinberg \& J.R.~Gott,
Phys. Rev. D {\bf 38} (1988) 2014.

\bibitem{l89}
G.~Lake,
Astron. J. {\bf 98} (1989) 1253.

\bibitem{Burkert:1997fz}
A.~Burkert,
in {\sl Aspects of Dark Matter in Astro and Particle Physics}, 
eds H.V. Klapdor \& Y. Ramachers (World Scientific, 1997) p.35. 

\bibitem{ss97}
P.~Salucci \& A.~Sinibaldi, 
Astron. Astrophys. {\bf 323} (1997) 1.

\bibitem{Sellwood:2000nh}
J.A.~Sellwood,
Astrophys. J. {\bf 540} (2000) L1.

\bibitem{Schramm:1980xv}
D.N.~Schramm \& G.~Steigman,
Astrophys. J.  {\bf 243} (1981) 1.

\bibitem{Blumenthal:1984bp}
G.R.~Blumenthal, S.M.~Faber, J.R.~Primack \& M.J.~Rees,
Nature {\bf 311} (1984) 517.

\bibitem{Padmanabhan:1993}
T.~Padmanabhan,
{\sl Structure Formation in the Universe} (Cambridge Univ Press, 1993).

\bibitem{Peebles:xt}
P.J.E.~Peebles,
{\sl Principles Of Physical Cosmology} (Princeton Univ Press, 1993).

\bibitem{Smoot:1992td}
G.F.~Smoot {\it et al.},
Astrophys. J.  {\bf 396} (1992) L1.

\bibitem{Doroshkevich:1980}
A.G.~Doroshkevich, Ya.B.~Zeldovich, R.A.~Sunyaev \& M.Yu.~Khlopov,
Sov. Astron. Lett. {\bf 6} (1980) 251, 265

\bibitem{Bond:ha}
J.R.~Bond, G.~Efstathiou \& J.~Silk,
Phys. Rev. Lett.  {\bf 45} (1980) 1980.

\bibitem{Doroshkevich:tq}
A.G.~Doroshkevich, M.Yu.~Khlopov, R.A.~Sunyaev, Ya.B.~Zeldovich \& A.S.~Szalay,
Ann. N.Y. Acad. Sci. {\bf 375} (1981) 32.

\bibitem{Peebles:ib}
P.J.E.~Peebles,
Astrophys. J.  {\bf 258} (1982) 415.

\bibitem{Klypin:1983}
A.A.~Klypin \& S.F.~Shandarin, 
Mon. Not. R. Astr. Soc. {\bf 204}\ (1983) 891.

\bibitem{White:yj}
S.D.M.~White, C.S.~Frenk \& M.~Davis,
Astrophys. J.  {\bf 274} (1983) L1.

\bibitem{Bond:hb}
J.R.~Bond \& A.S.~Szalay,
Astrophys. J.  {\bf 274} (1983) 443.

\bibitem{Peebles:1984}
P.J.E.~Peebles, 
Science {\bf 224} (1984) 1385.

\bibitem{Kaiser:1983}
N.~Kaiser, 
Astrophys. J. {\bf 273} (1983) L17.

\bibitem{White:84}
S.D.M.~White, M.~Davis \& C.~Frenk, 
Mon. Not. R. Astr. Soc. {\bf 209} (1984) 27p.

\bibitem{Zeng:1991}
Y.~Zeng \& S.D.M.~White, 
Astrophys. J. {\bf 374} (1991) 1.

\bibitem{Braun:1987uc}
E.~Braun, A.~Dekel \& P.R.~Shapiro,
Astrophys. J. {\bf 328} (1988) 34.

\bibitem{Centrella:1988}
J.~Centrella \etal, 
Astrophys. J. {\bf 333} (1988) 333.

\bibitem{Cen:1992}
R.~Cen \& J.P.~Ostriker, 
Astrophys. J. {\bf 399} (1992) 331.

\bibitem{Peebles:1982ff}
P.J.E.~Peebles,
Astrophys. J.  {\bf 263} (1982) L1.

\bibitem{Bond:fp}
J.R.~Bond \& G.~Efstathiou,
Astrophys. J.  {\bf 285} (1984) L45.

\bibitem{Davis:rj}
M.~Davis, G.~Efstathiou, C.S.~Frenk \& S.D.M.~White,
Astrophys. J.  {\bf 292} (1985) 371.

\bibitem{cdmrev}
J.P.~Ostriker, 
Ann. Rev. Astron. Astrophys. {\bf 31} (1993) 689.

\bibitem{Jungman:1995df}
G.~Jungman, M.~Kamionkowski \& K.~Griest,
Phys.\ Rept.\  {\bf 267} (1996) 195
[arXiv:hep-ph/9506380].

\bibitem{Linde:nc}
A.D.~Linde,
{\sl Particle Physics and Inflationary Cosmology} (Harwood Academic, 1990).

\bibitem{Liddle:cg}
A.R.~Liddle \& D.H.~Lyth,
{\sl Cosmological Inflation and Large-Scale Structure} 
(Cambridge Univ Press, 2000).

\bibitem{Sachs:er}
R.K.~Sachs \& A.M.~Wolfe,
Astrophys. J.  {\bf 147} (1967) 73.

\bibitem{Brandenberger:1987er}
R.H.~Brandenberger, N.~Kaiser, D.N.~Schramm \& N.~Turok,
Phys. Rev. Lett.  {\bf 59} (1987) 2371.

\bibitem{hdmdef}
E.~Bertschinger \& P.N.~Watts, 
Astrophys. J. {\bf 328} (1988) 23.

\bibitem{Gratsias:uc}
J.~Gratsias, R.J.~Scherrer, G.~Steigman \& J.V.~Villumsen,
Astrophys. J.  {\bf 405} (1993) 30.

\bibitem{hdmiso}
P.J.E.~Peebles, 
in {\sl The Origin and Evolution of Galaxies},
ed. B.~J.~T.~Jones (Reidel, 1983) p.143.

\bibitem{Sugiyama:1988qc}
N.~Sugiyama, M.~Sasaki \& K.~Tomita,
Astrophys. J.  {\bf 338} (1989) L45.

\bibitem{Wright:tf}
E.L.~Wright {\it et al.},
Astrophys. J.  {\bf 396} (1992) L13.

\bibitem{Davis:ui}
M.~Davis, F.J.~Summers \& D.~Schlegel,
Nature {\bf 359} (1992) 393.

\bibitem{Taylor:zh}
A.N.~Taylor \& M.~Rowan-Robinson,
Nature {\bf 359} (1992) 396.

\bibitem{Scott:1995uj}
D.~Scott, J.~Silk \& M.J.~White,
Science {\bf 268} (1995) 829.

\bibitem{cphdm}
Q.~Shafi \& F.W.~Stecker, 
Phys. Rev. Lett. {\bf 53} (1984) 1292.

\bibitem{Schaefer:ua}
R.K.~Schaefer, Q.~Shafi \& F.W.~Stecker,
Astrophys. J.  {\bf 347} (1989) 575.

\bibitem{cphdm2}
A. Van Dalen \& R.K. Schaefer, 
Astrophys. J. {\bf 398} (1992) 33.

\bibitem{Klypin:1992sf}
A.~Klypin, J.~Holtzman, J.~Primack \& E.~Regos,
Astrophys. J.  {\bf 416} (1993) 1.

\bibitem{Jing:xf}
Y.P.~Jing, H.J.~Mo, G.~Borner \& L.Z.~Fang,
Astron. Astrophys.  {\bf 284} (1994) 703.

\bibitem{Ma:xs}
C.P.~Ma \& E.~Bertschinger,
Astrophys. J.  {\bf 429} (1994) 22.

\bibitem{Pogosian:1994ns}
D.Y.~Pogosian \& A.A.~Starobinsky,
Astrophys. J.  {\bf 447} (1995) 465.

\bibitem{Liddle:1995ay}
A.R.~Liddle, D.H.~Lyth, R.K.~Schaefer, Q.~Shafi \& P.T.~Viana,
Mon. Not. Roy. Astron. Soc.  {\bf 281} (1996) 531.

\bibitem{Shafi:1996ha}
Q.~Shafi \& R.K.~Schaefer,
arXiv:hep-ph/9612478.

\bibitem{Ma:1994ub}
C.P.~Ma \& E.~Bertschinger,
Astrophys. J.  {\bf 434} (1994) L5.

\bibitem{Primack:1994pe}
J.R.~Primack, J.~Holtzman, A.~Klypin \& D.O.~Caldwell,
Phys. Rev. Lett.  {\bf 74}, 2160 (1995).

\bibitem{Athanassopoulos:1995iw}
C.~Athanassopoulos {\it et al.}  [LSND Collaboration],
Phys. Rev. Lett.  {\bf 75} (1995) 2650.

\bibitem{Fukuda:1994mc}
Y.~Fukuda {\it et al.} [Kamiokande Collaboration],
Phys. Lett. B {\bf 335} (1994) 237.

\bibitem{White:1995vk} 
M.J.~White, D.~Scott, J.~Silk \& M.~Davis,
Mon. Not. Roy. Astron. Soc.  {\bf 276} (1995) L69.

\bibitem{Ross:1995dq}
G.G.~Ross \& S.~Sarkar,
Nucl. Phys. B {\bf 461} (1996) 597.

\bibitem{Adams:1996yd}
J.A.~Adams, G.G.~Ross \& S.~Sarkar,
Phys. Lett. B {\bf 391} (1997) 271.

\bibitem{Riess:1998cb}
A.G.~Riess {\it et al.} [Supernova Search Team Collaboration],
Astron. J.  {\bf 116} (1998) 1009.

\bibitem{Perlmutter:1998np}
S.~Perlmutter {\it et al.}  [Supernova Cosmology Project Collaboration],
Astrophys. J.  {\bf 517} (1999) 565.

\bibitem{Bahcall:1999xn}
N.A.~Bahcall, J.P.~Ostriker, S.~Perlmutter \& P.J.~Steinhardt,
Science {\bf 284} (1999) 1481.

\bibitem{Sahni:1999gb}
V.~Sahni \& A.A.~Starobinsky,
Int. J. Mod. Phys. D {\bf 9} (2000) 373.

\bibitem{Elgaroy:2002bi}
O.~Elgaroy {\it et al.},
Phys. Rev. Lett. {\bf 89} (2002) 061301.

\bibitem{Abbott:1983mk}
L.F.~Abbott \& M.B.~Wise,
Phys. Lett. B {\bf 135} (1984) 279.

\bibitem{Hu:1995fq}
W.~Hu, D.~Scott, N.~Sugiyama \& M.J.~White,
Phys. Rev. D {\bf 52} (1995) 5498.

\bibitem{Bertschinger:1995er}
E.~Bertschinger,
arXiv:astro-ph/9506070.

\bibitem{Seljak:1996is}
U.~Seljak \& M.~Zaldarriaga,
Astrophys. J.  {\bf 469} (1996) 437.

\bibitem{Lyth:1996we}
D.H.~Lyth,
Phys. Rev. Lett. {\bf 78} (1997) 1861.

\bibitem{Hu:2001bc}
W.~Hu \& S.~Dodelson,
Ann. Rev. Astron. Astrophys. {\bf 40} (2002) 171.

\bibitem{Hu:1996qs}
W.~Hu, N.~Sugiyama \& J.~Silk,
Nature {\bf 386} (1997) 37.

\bibitem{deBernardis:2000gy}
P.~de Bernardis {\it et al.}  [Boomerang Collaboration],
Nature {\bf 404} (2000) 955.

\bibitem{Balbi:2000tg}
A.~Balbi {\it et al.},
Astrophys. J. {\bf 545} (2000) L1
[Erratum-ibid. {\bf 558} (2001) L145].

\bibitem{Dodelson:1995es}
S.~Dodelson, E.~Gates \& A.~Stebbins,
Astrophys. J.  {\bf 467} (1996) 10.

\bibitem{Wang:2001gy}
X.M.~Wang, M.~Tegmark \& M.~Zaldarriaga,
Phys. Rev. D {\bf 65} (2002) 123001.

\bibitem{Hannestad:2002xv}
S.~Hannestad,
Phys. Rev. D {\bf 66} (2002) 125011.

\bibitem{Hu:1997mj}
W.~Hu, D.J.~Eisenstein \& M.~Tegmark,
Phys. Rev. Lett.  {\bf 80} (1998) 5255.

\bibitem{Burles:2000zk}
S.~Burles, K.M.~Nollett \& M.S.~Turner,
Astrophys. J.  {\bf 552} (2001) L1.

\bibitem{Bennett:2003ba}
C.L.~Bennett {\it et al.},
Astrophys. J.  {\bf 583} (2003) 1.

\bibitem{Spergel:2003cb}
D.N.~Spergel {\it et al.},
arXiv:astro-ph/0302209.

\bibitem{Pierce:2003uh}
A.~Pierce \& H.~Murayama,
arXiv:hep-ph/0302131.

\bibitem{Giunti:2003cf}
C.~Giunti,
arXiv:hep-ph/0302173.

\bibitem{Bhattacharyya:2003gi}
G.~Bhattacharyya, H.~Pas, L.~Song \& T.J.~Weiler,
arXiv:hep-ph/0302191.

\bibitem{Klapdor-Kleingrothaus:md}
H.V.~Klapdor-Kleingrothaus, A.~Dietz \& I.V.~Krivosheina,
Found. Phys.  {\bf 32} (2002) 1181.

\bibitem{Dodelson:2001ux}
S.~Dodelson {\it et al.} [SDSS Collaboration],
Astrophys. J.  {\bf 572} (2001) 140.

\bibitem{ht64}
F. Hoyle \& R.J. Tayler, 
Nature {\bf 203} (1964) 1108.

\bibitem{p66}
P.J.E. Peebles, 
Phys. Rev. Lett. {\bf 16} (1966) 411.

\bibitem{Shvartsman:1969mm}
V.F.~Shvartsman,
Pisma Zh. Eksp. Teor. Fiz.  {\bf 9} (1969) 315
[JETP Lett.  {\bf 9} (1969) 184].

\bibitem{Steigman:kc}
G.~Steigman, D.N.~Schramm \& J.E.~Gunn,
Phys. Lett. B {\bf 66} (1977) 202.

\bibitem{Denegri:1989if}
D.~Denegri, B.~Sadoulet \& M.~Spiro,
Rev. Mod. Phys.  {\bf 62} (1990) 1.

\bibitem{Yang:gn}
J.M.~Yang, M.S.~Turner, G.~Steigman, D.N.~Schramm \& K.A.~Olive,
Astrophys. J.  {\bf 281} (1984) 493.

\bibitem{Steigman:1986nh}
G.~Steigman, K.A.~Olive, D.N.~Schramm \& M.S.~Turner,
Phys. Lett. B {\bf 176} (1986) 33.

\bibitem{Ellis:1985fp}
J.R.~Ellis, K.~Enqvist, D.V.~Nanopoulos \& S.~Sarkar,
Phys. Lett. B {\bf 167} (1986) 457.

\bibitem{Steigman:xp}
G.~Steigman, K.A.~Olive \& D.N.~Schramm,
Phys. Rev. Lett.  {\bf 43} (1979) 239.

\bibitem{Barger:2003zh}
V.~Barger, P.~Langacker \& H.S.~Lee,
arXiv:hep-ph/0302066.

\bibitem{Bernstein:ad}
J.~Bernstein, L.S.~Brown \& G.~Feinberg,
Rev. Mod. Phys.  {\bf 61} (1989) 25.

\bibitem{Sarkar:1995dd}
S.~Sarkar,
Rept. Prog. Phys.  {\bf 59} (1996) 1493.

\bibitem{Olive:1989xf}
K.A.~Olive, D.N.~Schramm, G.~Steigman \& T.P.~Walker,
Phys. Lett. B {\bf 236} (1990) 454.

\bibitem{Walker:ap}
T.P.~Walker, G.~Steigman, D.N.~Schramm, K.A.~Olive \& H.S.~Kang,
Astrophys. J.  {\bf 376} (1991) 51.

\bibitem{Copi:ev}
C.J.~Copi, D.N.~Schramm \& M.S.~Turner,
Science {\bf 267} (1995) 192.

\bibitem{Hata:1995tt}
N.~Hata, R.J.~Scherrer, G.~Steigman, D.~Thomas, T.P.~Walker, S.~Bludman \& P.~Langacker,
Phys. Rev. Lett.  {\bf 75} (1995) 3977.

\bibitem{Kernan:1996yz}
P.J.~Kernan \& S.~Sarkar,
Phys. Rev. D {\bf 54} (1996) 3681.

\bibitem{Sarkar:2002er}
S.~Sarkar,
arXiv:astro-ph/0205116.

\bibitem{fo} 
B.D. Fields \& K.A. Olive, 
Astrophys. J. {\bf 506} (1998) 177.

\bibitem{os} 
K.A. Olive \& E. Skillman
New Astron. {\bf 6} (2001) 119.

\bibitem{Sauer:2001vh}
D.~Sauer \& K.~Jedamzik,
arXiv:astro-ph/0104392.

\bibitem{izo} 
Y.I. Izotov \etal,
Astrophys. J. {\bf 527} (1999) 757.

\bibitem{ppr} 
M. Peimbert, A. Peimbert \& M.T. Ruiz,
Astrophys. J. {\bf 541} (2000) 688. 

\bibitem{ppl} 
A. Peimbert, M. Peimbert \& V. Luridiana,
Astrophys. J. {\bf 565} (2002) 668.

\bibitem{rnb} 
S.G. Ryan, J.E. Norris \& T.C. Beers,
Astrophys. J. {\bf 523} (1999) 654.

\bibitem{sfosw} 
E. Vangioni-Flam \etal, 
New Astron. {\bf 4} (1999) 245.

\bibitem{rbofn} 
S.G. Ryan, T.C. Beers, K.A. Olive, B.D. Fields \& J.E. Norris,
Astrophys. J. {\bf 530} (2000) L57.

\bibitem{vau} 
S. Vauclair \& C. Charbonnel,
Astrophys. J. {\bf 502} (1998) 372.

\bibitem{ddk} 
M.H. Pinsonneault, T.P. Walker, G. Steigman \& V.K. Narayanan,
Astrophys. J. {\bf 527} (1999) 180.

\bibitem{lin} 
J. Linsky, 
Space Sci. Rev. {\bf 84}, 285 (1998).

\bibitem{schr} 
A. Songaila, L.L. Cowie, C. Hogan \& M. Rugers, 
Nature {\bf 368} (1994) 599 .

\bibitem{webb}
J.K. Webb \etal, 
Nature {\bf 388} (1997) 250.

\bibitem{bkt}
S. Burles, D. Kirkman \& D. Tytler,
Astrophys. J. {\bf 519} (1999) 18.

\bibitem{Kirkman:2001nk}
D.~Kirkman {\it et al.},
Astrophys. J.  {\bf 559} (2001) 23.

\bibitem{ome} 
J.M. O'Meara \etal,
Astrophys. J. {\bf 552} (2001) 718.

\bibitem{Kirkman:2003uv}
D.~Kirkman, D.~Tytler, N.~Suzuki, J.M.~O'Meara \& D.~Lubin,
arXiv:astro-ph/0302006.

\bibitem{dod} 
S. D'Odorico, M. Dessauges-Zavadsky \& P. Molaro,
Astron. Astrophys. {\bf 368} (2001) L21.

\bibitem{pb} 
M. Pettini \& D. Bowen,
Astrophys. J. {\bf 560} (2001) 41. 

\bibitem{lev}
S.A. Levshakov, M. Dessauges-Zavadsky, S. D'Odorico \& P. Molaro,
Astrophys. J. {\bf 565} (2002) 696.

\bibitem{son} 
G. Sonneborn \etal, 
Astrophys. J. {\bf 545} (2000) 277.

\bibitem{ktblo} 
D. Kirkman, D. Tytler, S. Burles, D. Lubin \& J.M. O'Meara,
Astrophys. J. {\bf 529}(2000) 655. 

\bibitem{bbrw} 
T.M. Bania, R.T. Rood \& D.S. Balser,
Nature {\bf 415} (2002) 54.

\bibitem{dst} 
D. Galli, L. Stanghellini, M. Tosi \& F. Palla,
Astrophys. J. {\bf 477} (1997) 218.

\bibitem{char}
C. Charbonnel, 
Space Sci. Rev. {\bf 84}(1998) 199. 

\bibitem{Cyburt:2001pq}
R.H.~Cyburt, B.D.~Fields \& K.~A.~Olive,
Astropart. Phys.  {\bf 17} (2002) 87.

\bibitem{Adams:1997de}
J.A.~Adams, G.G.~Ross \& S.~Sarkar,
Nucl. Phys. B {\bf 503} (1997) 405.

\bibitem{Barriga:2000nk}
J.~Barriga, E.~Gaztanaga, M.G.~Santos \& S.~Sarkar,
Mon. Not. Roy. Astron. Soc.  {\bf 324} (2001) 977.

\bibitem{Bridle:2003sa}
S.L.~Bridle, A.M.~Lewis, J.~Weller \& G.~Efstathiou,
arXiv:astro-ph/0302306.

\bibitem{Lisi:1999ng}
E.~Lisi, S.~Sarkar \& F.L.~Villante,
Phys. Rev. D {\bf 59} (1999) 123520.

\bibitem{Burles:1999zt}
S.~Burles, K.M.~Nollett, J.N.~Truran \& M.S.~Turner,
Phys. Rev. Lett.  {\bf 82} (1999) 4176.

\bibitem{DiBari:2001ua}
P.~Di Bari,
Phys. Rev. D {\bf 65} (2002) 043509.

\bibitem{Abazajian:2002bj}
K.~N.~Abazajian,
arXiv:astro-ph/0205238.

\bibitem{Hannestad:2003xa}
S.~Hannestad,
arXiv:astro-ph/0302340.

\bibitem{Cyburt:2003fe}
R.H.~Cyburt, B.D.~Fields \& K.A.~Olive,
arXiv:astro-ph/0302431.

\bibitem{Pierpaoli:2003}
E. Pierpaoli,
arXiv:astro-ph/0302465.

\bibitem{Crotty:2003th}
P.~Crotty, J.~Lesgourgues \& S.~Pastor,
arXiv:astro-ph/0302337.

\bibitem{degen}
H.-S. Kang \& G. Steigman, 
Nucl. Phys. {\bf B372}, 494 (1992).

\bibitem{Lunardini:2000fy}
C.~Lunardini \& A.Y.~Smirnov,
Phys. Rev. D {\bf 64} (2001) 073006.

\bibitem{Dolgov:2002ab}
A.D.~Dolgov, S.H.~Hansen, S.~Pastor, S.T.~Petcov, G.G.~Raffelt \& D.V.~Semikoz,
Nucl. Phys. B {\bf 632} (2002) 363.

\bibitem{Wong:2002fa}
Y.Y.~Wong,
Phys. Rev. D {\bf 66} (2002) 025015.

\bibitem{Abazajian:2002qx}
K.N.~Abazajian, J.F.~Beacom \& N.F.~Bell,
Phys. Rev. D {\bf 66} (2002) 013008.

\bibitem{Applegate:hm}
J.H.~Applegate, C.J.~Hogan \& R.J.~Scherrer,
Phys. Rev. D {\bf 35} (1987) 1151.

\bibitem{inhom2}
K. Kainulainen, H. Kurki-Suonio \& E. Sihvola,
Phys. Rev. {\bf D59}, 083505 (1999);\\
K. Jedamzik \& J.B. Rehm, 
Phys. Rev. {\bf D64}, 023510 (2001).

\bibitem{antimat}
J.B. Rehm \& K. Jedamzik,
Phys. Rev. Lett. {\bf 81}, 3307 (1998);\\
H. Kurki-Suonio \& E. Sihvola,
Phys. Rev. {\bf D62}, 103508 (200).

\bibitem{desh}
S. Dimopoulos, R. Esmailzadeh, G.D. Starkman \& L.J. Hall,
Astrophys. J. {\bf 330} (1988) 545;\\
K. Jedamzik,
Phys. Rev. Lett. {\bf 84} (2000) 3248.

\bibitem{Arkani-Hamed:1998vp}
N.~Arkani-Hamed, S.~Dimopoulos, G.R.~Dvali \& J.~March-Russell,
Phys. Rev. D {\bf 65} (2002) 024032.

\bibitem{Abazajian:2000hw}
K.~Abazajian, G.M.~Fuller \& M.~Patel,
arXiv:hep-ph/0011048.

\end{thebibliography}
\end{document}